\documentclass[aps, prl,twocolumn,superscriptaddress]{revtex4-1}

\usepackage[table]{xcolor}
\usepackage{dcolumn}% Align table columns on decimal point
\usepackage{bm}% bold math
\usepackage[T1]{fontenc}
\usepackage{braket}
\usepackage{amsmath}
\usepackage{amssymb}
\usepackage{nccmath}
\usepackage{mathrsfs}
\usepackage{multirow}
\usepackage{ulem}

\usepackage{pgf}   % need for figures
\usepackage{graphicx}
\let\oldsim\sim 
\renewcommand{\sim}{{\oldsim}}

\graphicspath{ {./images/} }

\begin{document}
	\title{Magnetic trapping of ultracold molecules at high density}

\author{Juliana J. Park}
\affiliation{Research Laboratory of Electronics, MIT-Harvard Center for Ultracold Atoms,
Department of Physics, Massachusetts Institute of Technology, Cambridge, Massachusetts 02139, USA}

\author{Yu-Kun Lu}
\affiliation{Research Laboratory of Electronics, MIT-Harvard Center for Ultracold Atoms,
Department of Physics, Massachusetts Institute of Technology, Cambridge, Massachusetts 02139, USA}

\author{Alan O. Jamison}
\affiliation{Institute for Quantum Computing and Department of Physics \& Astronomy,
University of Waterloo, Waterloo, Ontario N2L 3G1, Canada}

\author{Wolfgang Ketterle}
\affiliation{Research Laboratory of Electronics, MIT-Harvard Center for Ultracold Atoms,
Department of Physics, Massachusetts Institute of Technology, Cambridge, Massachusetts 02139, USA}

%\wk{[I would first say what we have achieved and not discuss the comparison with optical traps in the first or second sentence]} Trapping of atoms and molecules in a conservative trap is essential for enhancement in their densities or phase-space densities for their applications, including quantum state-controlled chemistry, quantum simulations, and quantum information processing. Although optical traps have the advantage of providing tight confinement to magnetic traps, which has allowed the study of molecular collisions, they may cause light-assisted collisional loss in molecular systems. We report magnetic trapping of NaLi molecules in the triplet ground state at high density ($\approx 10^{11} \; \rm{cm}^{-3}$) and low temperature ($\approx 1\;{\rm \mu K}$) allowing the observation of molecular collisions in the ultracold regime. Studies on both atom-molecule and molecule-molecule collisions are possible in this regime. We measure the inelastic loss rates in a single spin sample and spin-mixtures of a fermionic NaLi molecular gas and spin-stretched NaLi$+$Na mixture. We demonstrate sympathetic cooling of NaLi molecules in the magnetic trap by radio frequency evaporation of Na atoms that are co-trapped with molecules, and observe an increase in the phase space density by a factor of $\approx 16$.
	\begin{abstract}
Trapping ultracold molecules in conservative traps is essential for applications---such as quantum state-controlled chemistry, quantum simulations, and quantum information processing. These applications require high densities or phase-space densities. We report magnetic trapping of NaLi molecules in the triplet ground state at high density ($\approx 10^{11} \; \rm{cm}^{-3}$) and ultralow temperature ($\approx 1\;{\rm \mu K}$). Magnetic trapping at these densities allows studies on both atom-molecule and molecule-molecule collisions in the ultracold regime in the absence of trapping light, which has often lead to undesired photo-chemistry. We measure the inelastic loss rates in a single spin sample and spin-mixtures of fermionic NaLi as well as spin-stretched NaLi$+$Na mixtures. We demonstrate sympathetic cooling of NaLi molecules in the magnetic trap by radio frequency evaporation of co-trapped Na atoms and observe an increase in the molecules' phase-space density by a factor of $\approx 16$.
	\end{abstract}
	\maketitle

\section{Introduction}

Ultracold molecules offer new opportunities for quantum state controlled chemistry \cite{krems2008cold, balakrishnan2016perspective}, for quantum simulations \cite{micheli2006toolbox, capogrosso2010quantum, blackmore2018ultracold}, and for quantum information processing \cite{ni2018dipolar, herrera2014infrared, hughes2020robust, sawant2020ultracold}. 
For more than two decades, various methods were developed with the goal to trap molecular samples at densities high enough to study molecular collisions.  Using buffer gas cooling \cite{weinstein1998magnetic, campbell2007magnetic, lu2014magnetic, tsikata2010magnetic} or Stark or magnetic deceleration \cite{liu2015one,liu2017magnetic, riedel2011accumulation},  molecules were trapped in magnetic traps in the temperature range of tens or hundreds of millikelvin. In only one case ($\rm{O}_2$) were densities high enough (estimated at $\approx 10^{10}\;\rm{cm}^{-3}$) to observe molecular collisions \cite{segev2019collisions}.  Dipolar molecules can be trapped with electric forces.  Collisions between CH$_3$F molecules were observed in an electrostatic trap at densities of $10^{7}\;\rm{cm}^{-3}$ \cite{koller2022electric}, and elastic collisions of OH were observed in an electromagnetic trap at temperatures around 60 mK \cite{reens2017controlling}, though none of these studies reached the ultracold temperature regime.  
Ultracold laser-cooled molecules (in the 50 - 100 $\mu$K range) were transferred into magnetic traps at densities $\lesssim 10^{6}\;\rm{cm}^{-3}$ \cite{mccarron2018magnetic, williams2018magnetic}, too low for the observation of intermolecular collisions.

Optical traps have the advantages of trapping non-magnetic molecules and providing tight confinement, which has allowed the study of collisions involving molecules created by ultracold assembly or direct laser cooling. For example, collisional resonances have been observed in atom-molecule mixtures \cite{son2022control,yang2019observation, wang2021magnetic} and between molecules \cite{park2022feshbach} in optical traps. 
%\cite{hudson2008inelastic, ni2010dipolar, ospelkaus2010quantum, de2011controlling, takekoshi2014ultracold, drews2017inelastic, ye2018collisions, gregory2019sticky} or direct laser cooling \cite{cheuk2020observation}.  
The disadvantage of optical traps is the small trapping volume and the presence of intense laser light, which can induce photochemistry.  This has become a major concern recently after many experiments have found fast collisional losses even for non-reactive molecules, possibly due to ``sticky collisions'' connected with long-lived complexes  \cite{mayle2012statistical, mayle2013scattering, christianen2019quasiclassical, christianen2019photoinduced, liu2022bimolecular}. Recent experiments to test these proposals suggest that optical traps can cause short lifetimes of molecules and are not truly conservative \cite{liu2020photo, gregory2020loss, gregory2021molecule, ComplexesNaKNaRb, bause2021collisions} emphasizing the need for ``laser-free'' trapping.
%\cite{krb_kkni, narb, caf, rb2, li2, cs2cs, cs2cs_2, liscscs, rbcscs_higher_similar}

Here, we report magnetic trapping of triplet NaLi molecules in the rovibrational ground state with high density ($\approx 10^{11} \; \rm{cm}^{-3}$) and ultracold temperature ($\approx 1 \mu$K). The typical density is a factor of $10^5$ higher compared to previous experiments with magnetically trapped ultracold molecular gases. Inelastic losses are detected in single-component and in spin-mixtures of a fermionic NaLi molecular gas.

Another major long-standing goal has been magnetic trapping of molecules together with atoms for sympathetic cooling of molecules to achieve higher molecular densities or phase-space densities \cite{lara2006ultracold, zuchowski2008prospects}. Magnetic co-trapping of NH and N \cite{hummon2011cold}, and more recently of CaF and Rb \cite{jurgilas2021collisions} and $\rm{O}_2$ with Li atoms \cite{akerman2017trapping} has been achieved.  However, so far only inelastic collisions were observed \cite{hummon2011cold, jurgilas2021collisions}, with atomic densities far too low for sympathetic cooling.

Here we demonstrate sympathetic cooling of molecules in a magnetic trap.  We use a spin-stretched NaLi$+$Na mixture, which had been studied in ref. \cite{NaLiSympCool} and observe an increase in the phase-space density (PSD) of the molecular gas by more than an order of magnitude after radio frequency (RF) evaporation of Na atoms.

\section{Experimental protocol $\&$ results}

\begin{figure*}  
    \centering  
	\includegraphics[width=170mm,keepaspectratio]{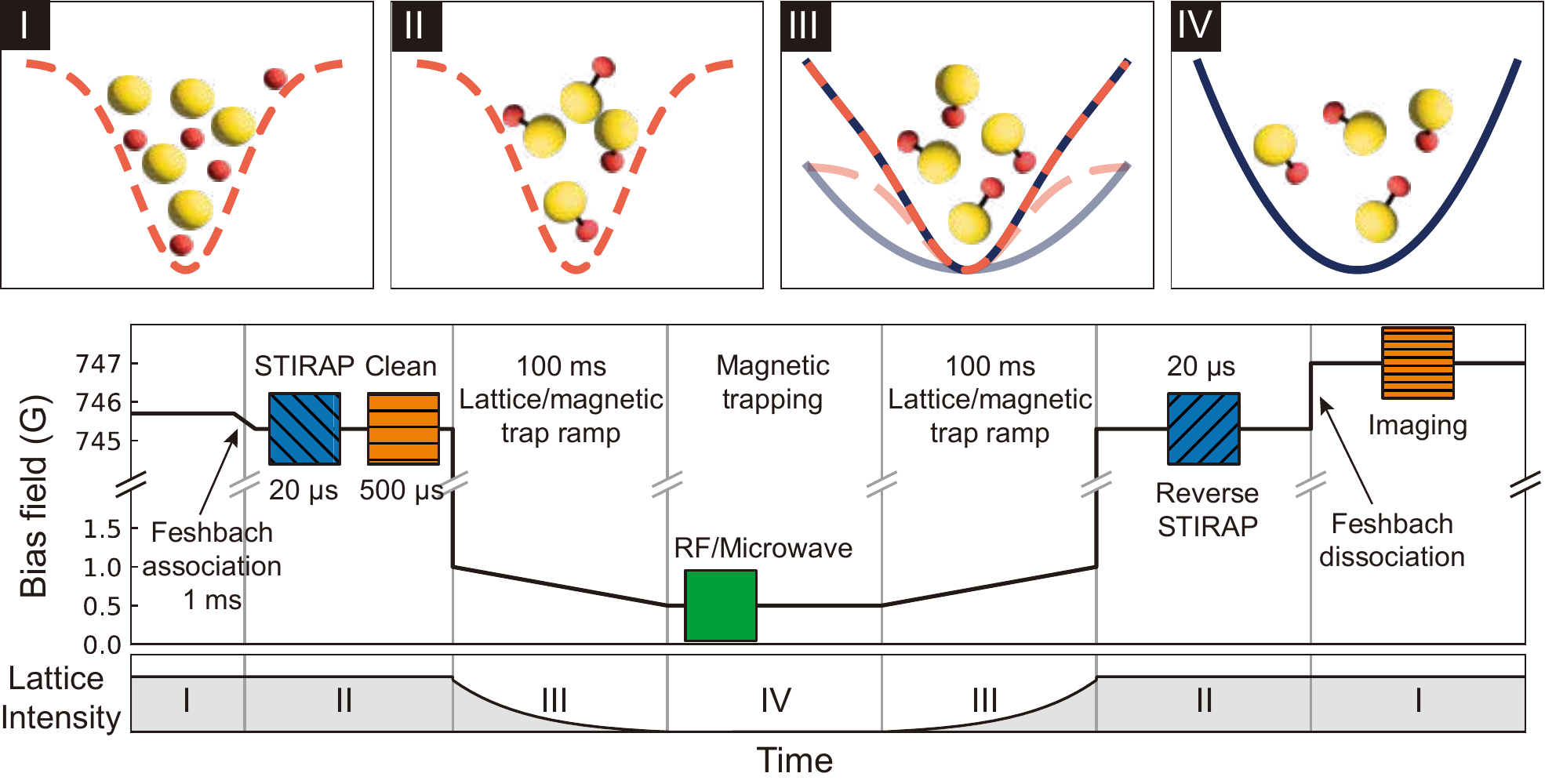}
	\caption{Top figures are the illustration of particles confined in different traps. I. Na/Li mixture is confined in a 1D lattice potential (indicated with orange dashed line). II. NaLi molecules are trapped in the same 1D lattice. III. NaLi molecules are trapped in a hybrid trap created by a 1D lattice and a magnetic trap (indicated with black line). IV. Molecules are purely confined by a magnetic trap. The middle plot is the experimental sequence to produce and isolate NaLi molecules (time axis is not to scale). Molecules are formed via Feshbach formation and STIRAP in a 1D lattice, and the free atoms are removed using resonant light pulses at high field. After the magnetic bias field is dropped to low field, the lattice is ramped down and the magnetic trap is turned on in 100 ms. In the magnetic trap RF or microwave is applied to the molecules for thermometry or preparation of molecular spin mixtures. For detection, molecules are transferred back to the 1D lattice, the molecule formation process is reversed, and the dissociated free atoms are imaged. The bottom row is the lattice intensity as a function of time along with particle and trap type indicated using index I-IV. }
	\label{fig:experimental sequence}
\end{figure*}

\begin{figure*}  
    \centering  
	\includegraphics[width=170mm,keepaspectratio]{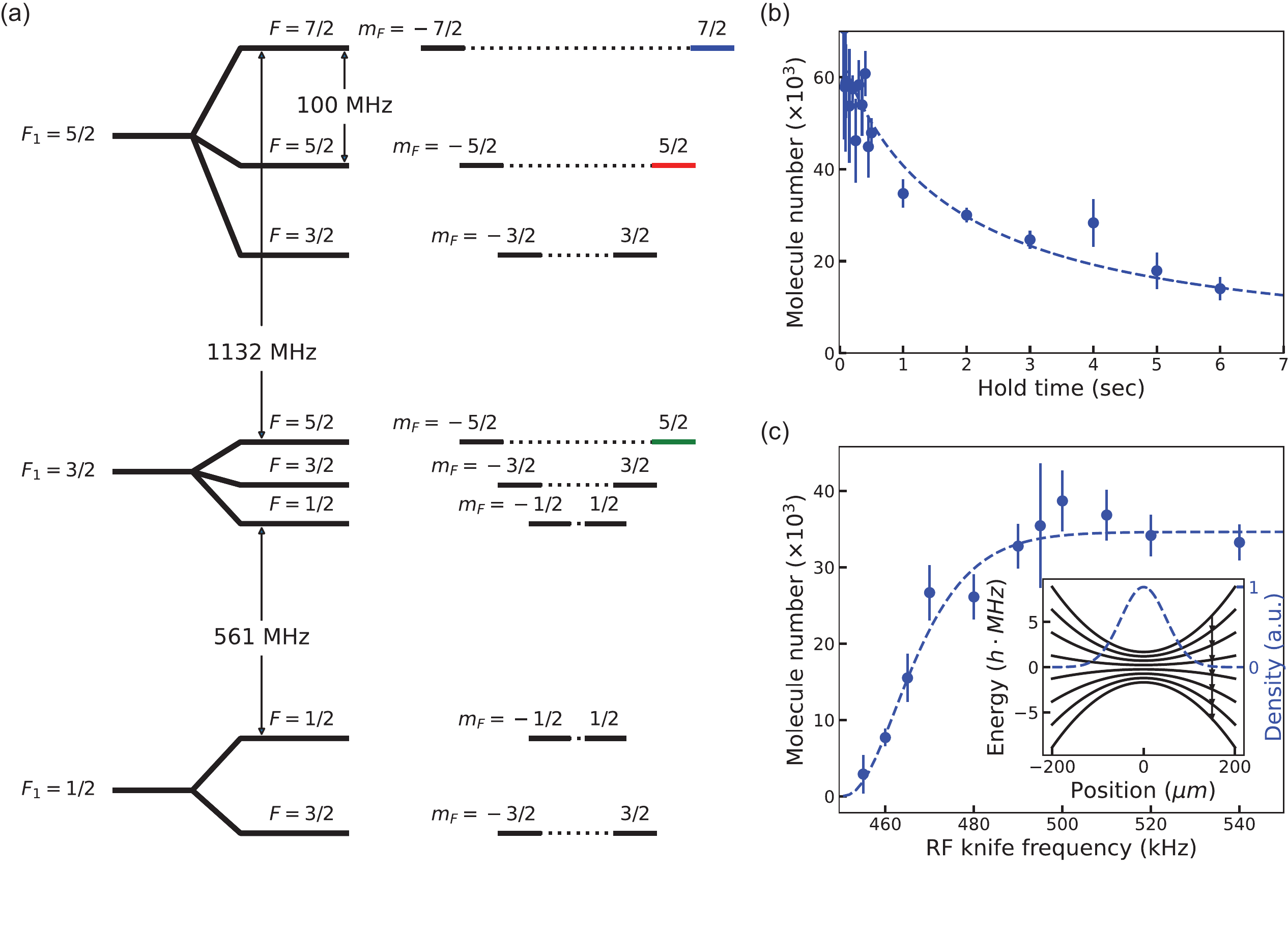}
	\caption{(a) Rovibrational ground state energy level diagram of NaLi molecules in the triplet potential ($a^{3}\Sigma^{+}$) at zero bias field. Quantum number $ F = \lvert \vec{F} \rvert = \lvert \vec{S} + \vec{I}_{\rm{Na}} + \vec{I}_{\rm{Li}}\rvert$ is the total angular momentum, where $S = \lvert \vec{S} \rvert = 1$ is the total electron spin of NaLi and $I_{\text{Na}} = 3/2$ and $I_{\rm{Li}} = 1$ are the nuclear spins of Na and Li. In the zero field limit, $F_1 = \lvert \vec{S} + \vec{I}_{\rm{Na}} \rvert = 1/2, 3/2$, and $5/2$ is an approximately good quantum number that characterizes the largest-scale hyperfine splittings as the hyperfine splitting due to the Na nucleus is significantly larger than that due to the Li nucleus. Molecules are initially formed in the low field seeking spin-polarized hyperfine state ($\ket{F_{1}=5/2, F=7/2, m_{F}=7/2}$) indicated in blue. States in red and green are the other hyperfine states used for creating molecular spin-mixtures. (b) Number decay of low field seeking spin-polarized NaLi molecules ($\ket{F_{1}=5/2, F=7/2, m_{F}=7/2}$) in a magnetic trap. (c) The number of molecules left in the magnetic trap as a function of the RF-knife frequency. The dashed line is a fitting function for a temperature of 1.06 $\mu$K.  The inset is the $m_F$ energy levels of NaLi in $\ket{ F_1=5/2, F=7/2}$ near the center of a magnetic trap (in black lines). Exemplary density profile of molecules in the top hyperfine state is in blue dashed line.}
	\label{fig:combined}
\end{figure*}

The experiments are carried out with NaLi molecules in the triplet ground state ($a^{3}\Sigma^{+}$, $v=0$, $N=0$) created by means of ultracold assembly of Na and Li atoms in the lowest-energy Zeeman states 
%(in the $\ket{F,m_{F}}_{\rm{atom}}$ basis, $\ket{1,1}$ for Na, and $\ket{1/2,1/2}$ for Li) 
following the procedure described in ref. \cite{NaLiSympCool, son2022control, park2022feshbach}. The procedure has been improved so that the Na/Li mixture is transferred directly from the initial magnetic trap into a $1550\;{\rm nm}$ 1D optical lattice without the use of extra optical traps (previous configurations required a $1064\;{\rm nm}$ optical dipole trap and a $1596\;{\rm nm}$ 1D optical lattice). After 0.4 seconds of forced evaporation of the atom mixture, $10^5$ molecules at 1.8 $\mu$K temperature are formed in the maximally stretched low-field seeking hyperfine state ($\ket{F_1=5/2, F=7/2, m_{F}=7/2}$) using Feshbach association and STIRAP transfer to the triplet ground state. Here, $ F $ is the total angular momentum including electron and nuclear spins, $m_F$ is the $F$ projection to the quantization axis, and $F_1$ is a good quantum number in the zero-field limit combining the electron spin and the Na nuclear spin \cite{NaLiGround}.
%Here, $ F = \lvert \vec{F} \rvert = \lvert \vec{S} + \vec{I}_{\rm{Na}} + \vec{I}_{\rm{Li}}\rvert$ is the total angular momentum, where $S = \lvert \vec{S} \rvert = 1$ is the total electron spin of NaLi and $I_{\text{Na}} = 3/2$ and $I_{\rm{Li}} = 1$ are the nuclear spins of Na and Li. In the zero field limit, $F_1 = \lvert \vec{S} + \vec{I}_{\rm{Na}} \rvert = 1/2, 3/2$, and $5/2$ is an approximately good quantum number that characterizes the largest-scale hyperfine splittings as the hyperfine splitting due to the Na nucleus is significantly larger than that due to the Li nucleus. The energy level diagram of NaLi molecules in the triplet ground state ($a^{3}\Sigma^{+}$, $v=0$, $N=0$) is shown in Fig. \ref{fig:combined}(a) 

After ramping down the bias field from $\sim$ 745 G, where the association of molecules occurs, to a low magnetic bias field in 20 ms, molecules are transferred from the optical lattice to an Ioffe-Pritchard magnetic trap with a bias field of $0.56\;{\rm G}$ in 100 ms. The molecules are trapped for various hold times and transferred back to the optical lattice in 100 ms for detection at high field ($\sim 745$ G). The number of molecules is counted by absorption imaging of the Li atoms from the recaptured and dissociated molecules. The experimental sequence is illustrated in Figure \ref{fig:experimental sequence}. The transfer efficiency of molecules from the optical trap to the magnetic trap is close to 100 \% whereas the recapture efficiency back to the optical trap is only about 50 \% due to a smaller optical trap volume.

\begin{figure}  
    \centering  
	\includegraphics[width=83mm,keepaspectratio]{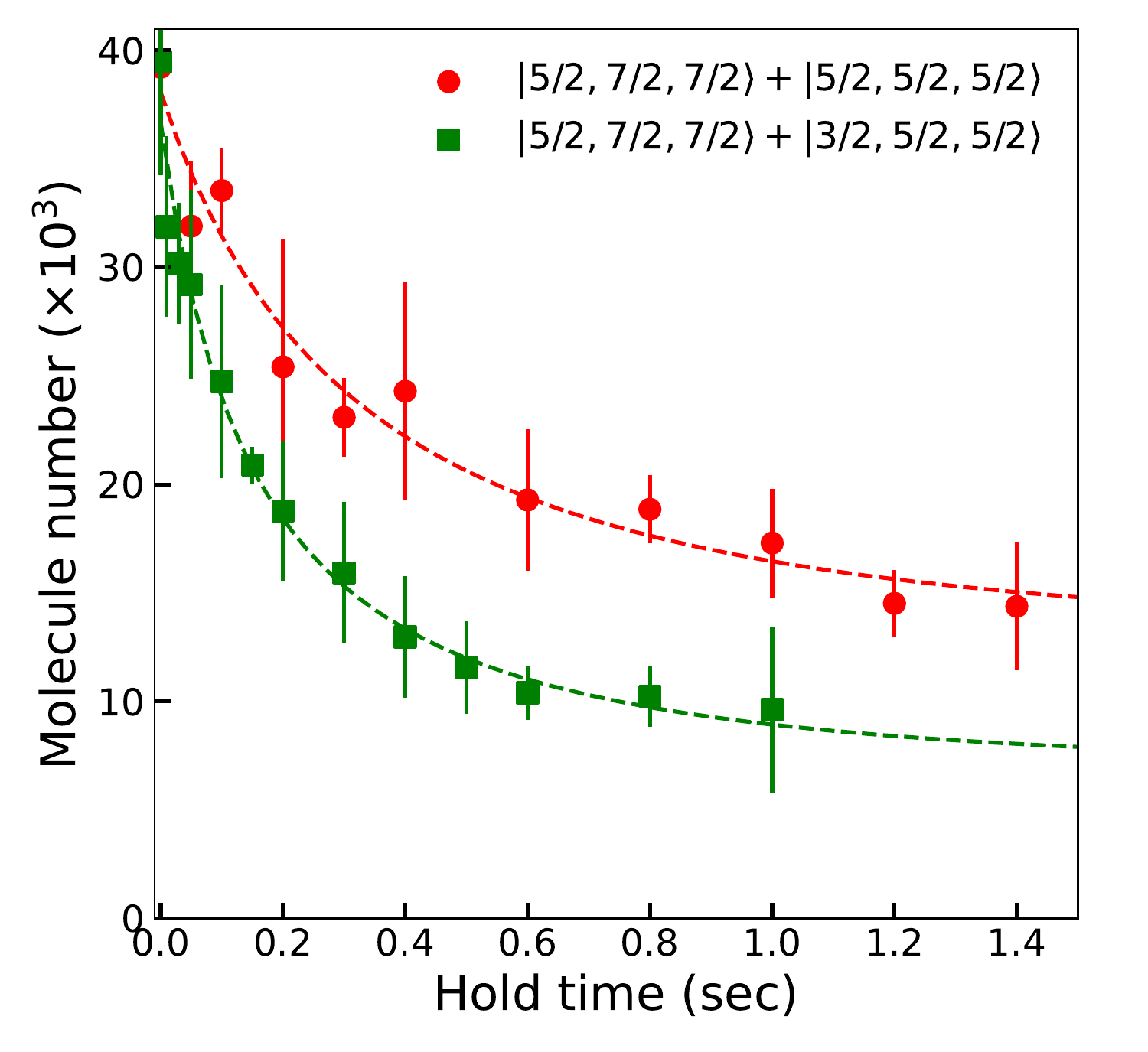}
	\caption{Number decay of low field seeking spin-polarized NaLi molecules ($\ket{F_{1}=5/2, F=7/2, m_{F}=7/2}$) when magnetically trapped together with molecules in state $\ket{F_{1}=5/2, F=5/2, m_{F}=5/2}$ (in red circles) and with molecules in state $\ket{F_{1}=3/2, F=5/2, m_{F} =5/2}$ (in green squares).}
	\label{fig:mol mixture}
\end{figure}

We investigate mainly three types of inelastic collisions: $p$-wave collisions in single spin component molecular gas, $s$-wave collisions in molecular spin-mixutures, and low-reactivity atom-molecule collisions i.e. NaLi$+$Na in the low field seeking spin-stretched states. For this, molecules in the hyperfine state $\ket{F_{1}=5/2, F=7/2, m_{F}=7/2}$ are held by the magnetic trap in the single spin state, or together with molecules in another hyperfine state or Na atoms in the upper spin-stretched state ($\ket{F=2, m_F=2}$). The energy level diagram of NaLi molecules in the triplet ground state is shown in Fig. \ref{fig:combined}(a). We focus on the number decay of NaLi in state $\ket{F_{1}=5/2, F=7/2, m_{F}=7/2}$ in the magnetic trap.

Densities are obtained from an estimation of the magnetic trap volume. For this, we measure the trap frequencies for Na atoms in the upper spin-stretched state ($\ket{F=2, m_F=2}$) and the temperature of the upper stretched NaLi or Na atoms (see Methods for details). 
The measured Na trap frequencies are $(f_x, f_y, f_z){=} (282, 282, 14.8)$ Hz.
For the thermometry of NaLi, we let the spin-polarized NaLi molecules in the $\ket{F_{1}=5/2, F=7/2, m_{F}=7/2}$ hyperfine state undergo RF-induced spin-flips in the magnetic trap to map the density distribution of the molecules. RF is swept from some high value ($ > 600$ kHz) to various final RF values (RF-knife frequencies) $f$ in 600 ms (faster than collisional rethermalization timescales). This RF induces 7 simultaneous resonant spin-flips from $\ket{F_{1}=5/2, F=7/2, m_{F}=7/2}$ to $\ket{F_{1}=5/2, F=7/2, m_{F}=-7/2}$, allowing the molecules to escape the trap as depicted in the inset of Fig. \ref{fig:combined}(c). The number of molecules remaining in the magnetic trap as a function of the RF-knife frequency determines the density distribution of the molecular gas in the magnetic trap and therefore the temperature as well. The best estimate for the temperature of the other molecular spin component (indicated with NaLi$^{*}$) is given by $T_{\text{NaLi}^{*}} \approx T_{\text{NaLi}}( \mu_{\text{NaLi}^{*}}/\mu_{\text{NaLi}}+1)/2$, where $T_{\rm{NaLi}}$ is the temperature of NaLi in the upper stretched state, and $\mu_{\text{NaLi}^{*}}/\mu_{\text{NaLi}}$ is the magnetic moment ratio (see Methods for details).
The Na temperature is measured directly from the time-of-flight (TOF) absorption imaging out of the magnetic trap at low field. 

%The experimental data in Fig. \ref{fig:combined}(c) provide an estimate of the molecule temperature which is 1.05 uK . 

We first observe $p$-wave inelastic collisions of NaLi in the upper spin-stretched state. The number of molecules in the magnetic trap decays by more than 50\% starting from $6\times 10^{4}$ in a few seconds, as shown in Fig. \ref{fig:combined}(b), while the typical vacuum lifetime is greater than 20 seconds. 
%Estimation of the magnetic trap geometry and the molecular temperature in the magnetic trap are crucial in obtaining the density and loss rate constant. The magnetic trap frequencies of the upper spin-stretched Na ($\ket{F=2, m_F=2}$) were measured to be $(f_x, f_y, f_z){=} (281, 281, 14.8)$ Hz. For thermometry, we let the spin-polarized NaLi molecules in the $\ket{F_{1}=5/2, F=7/2, m_{F}=7/2}$ hyperfine state undergo rf-induced spin-flips in the magnetic trap to map the density distribution of the molecule gas. rf is swept from some high value($ > 600kHz$) to a lower value $f$ in 600 ms. This rf can induce 7 simultaneous spin-flips by one quanta of $m_F$ respectively from state $\ket{F_{1}=5/2, F=7/2, m_{F}=7/2}$ to $\ket{F_{1}=5/2, F=7/2, m_{F}=-7/2}$, allowing the molecules to escape the trap as depicted in the inset of Fig. \ref{fig:combined}(c). The number of molecules remaining in the magnetic trap is determined by the final rf knife frequency, since the maximum possible kinetic energy of the molecules that are trapped is given by $A=\frac{7}{2} h (f-f_0)$ where $f_0$ is the transition frequency from $\ket{F_{1}=5/2, F=7/2, m_{F}=7/2}$ to $\ket{F_{1}=5/2, F=7/2, m_{F}=5/2}$ at the bottom of the magnetic trap. The experimental data in Fig. \ref{fig:combined}(c) provide an estimate of the molecule temperature which is 1.05 uK (see Methods for details).
The molecular temperature is $1.06\pm 0.18 \; \mu$K, which is measured by applying an RF-knife of various frequencies as shown in Fig. \ref{fig:combined}(c).  The initial density is $1.7 \times 10^{11} \; \rm{cm}^{-3}$---a factor of $10^5$ greater than the experiments carried out in ref. \cite{mccarron2018magnetic, williams2018magnetic}. The loss rate coefficient is calculated as $(3.6 \pm 1.4) \times 10^{-12}\;  (\rm{cm}^{3} /s) (T_{\rm{NaLi}}/\mu \rm{K})$ from the best fit to a two-body loss model described in \cite{park2022feshbach} (see Methods for details). Within the uncertainty, this value is consistent with the value reported in ref.\cite{son2022control}, which was measured near 980 G. 
%a factor of 2 larger than the $p$-wave universal value, $K^{univ}_{l=1}/T=(1.2 \pm 0.3) \times 10^{-12}\;{\rm cm^{3}/s \cdot \mu K}$, which is estimated using an approximate value of the NaLi-NaLi long-range dispersion coefficient ($C_{6} =5879$ a.u) obtained by summing all $C_{6}$ coefficients between the two constituent atoms \cite{derevianko2001high}. 

%The molecular temperature is $1.05\pm 0.18 \; \mu$ K, which is measured by applying an RF-knife of various frequencies and 
%The universal loss rate constant for $p$-wave ($s$-wave) collisions is $K^{univ}_{l=1}/T=(1.2 \pm 0.3) \times 10^{-12}\;{\rm cm^{3}/s \cdot \mu K}$ ($K^{univ}_{l=0}=1.85 \times 10^{-10}\; {\rm cm^{3}/s}$). 

\begin{figure}  
    \centering  
	\includegraphics[width=83mm,keepaspectratio]{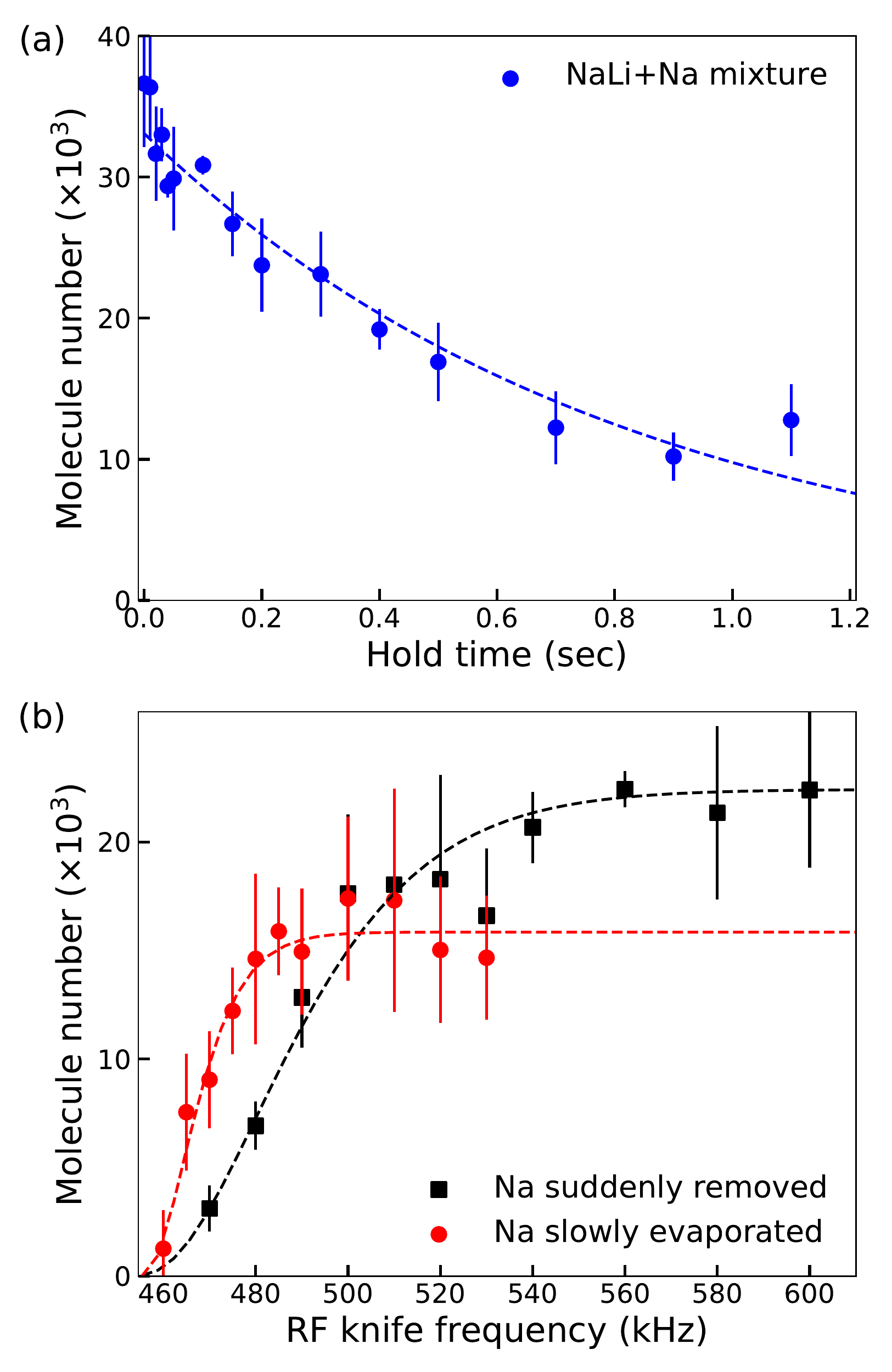}
	\caption{(a) Molecule number as a function of hold time in a magnetic trap when trapped together with Na. The atom-molecule mixture is in the low-field seeking spin-stretched state. Na number is about $4 \times 10^{5}$ and the temperature is $\approx 2 \mu$K. (b) Molecule number as a function of RF-knife frequency. Data in red circles is with slow evaporation of Na and data in black squares is with Na suddenly removed after loading into a magnetic trap. The density profile obtained from the red (black) data implies a temperature of 0.8(1) $\mu$K
	( $ 2.3(3) \;\mu$K). The fits in dashed lines use a fit function described in the Methods section.}
	\label{fig:atom and molecule}
\end{figure}

Next, we observe $s$-wave inelastic collisions by creating a spin-mixture of molecules in two different hyperfine states. A transition from $\ket{F_1=5/2, F=7/2,m_{F}=7/2}$ to $\ket{F_1=5/2, F=5/2,m_{F}=5/2}$ is driven by 100 MHz RF until the molecules form a near 50/50 spin mixture in the magnetic trap. Similarly, the mixture of $\ket{5/2, 7/2,7/2}$ and $\ket{3/2, 5/2, 5/2}$ is prepared with a 1133 MHz microwave drive. Due to magnetic field inhomogeneity, the superposition of the two spin states decoheres within $T_2 \approx \frac{h}{\Delta\mu \cdot \delta B} \lesssim 500 \mu \rm{s}$, where $\Delta\mu$ is the magnetic moment difference between the two hyperfine states, and $\delta B \approx 10 \; \rm{mG}$ is the magnetic field inhomogeneity, during and after an approximate $\pi/2$ pulse. The loss rate coefficients are calculated to be $(3.6 \pm 1.5) \times 10^{-11} \; \rm{cm}^{3}/s$ and $(1.1\pm 0.4) \times 10^{-10} \; \rm{cm}^{3}/s$ respectively by fitting the data in Fig. \ref{fig:mol mixture} to a loss model described in the Methods section. These values are significantly lower than the $s$-wave universal value estimated using the approximate NaLi-NaLi long-range dispersion coefficient ($K^{univ}_{l=0}=1.85 \times 10^{-10}\; {\rm cm^{3}/s}$).

For atom-molecule collisions, we load together about $4\times10^{5}$ Na atoms and about $4 \times10^{4}$ NaLi in the upper stretched hyperfine states into the magnetic trap. The temperature of the atoms is $\approx 2 \mu$K from the TOF imaging. Since the Na number is much greater than the NaLi number we fit the molecule decay in Fig. \ref{fig:atom and molecule}(a) to a simple exponential decay curve. 
The measured loss rate constant $\beta=(6.3 \pm 1.4) \times 10^{-12} \rm{cm}^{3}/s$ is about a factor of 30 lower than the universal value ($1.72\times 10^{-10} \rm{cm}^{3}/s$) which is well known for this system \cite{hermsmeier2021quantum}. 

%similar to ref.\cite{NaLiSympCool} but

With the low-reactivity atom-molecule mixture, we demonstrate sympathetic cooling of molecules in the magnetic trap.  Here, via RF controlled evaporation, we have independent control over the molecule and atom trap depths. 
It is possible to cool spin-polarized NaLi using collisions with Na, also spin-polarized in the same direction as NaLi, because NaLi has a favorable ratio of elastic to inelastic collisions $\gamma$ with Na. At high field, it was measured to be $\gamma \approx 300$ \cite{NaLiSympCool}.
After loading the atom-molecule mixture into the magnetic trap, we slowly evaporate Na atoms out of the trap with a microwave sweep. %Since the temperature of NaLi in a magnetic trap cannot be measured by time of flight imaging but requires molecule thermometry using the RF-knife scan as in Fig. \ref{fig:combined}(c), which is achieved by more than 30 experimental runs, we did not attempt to optimize the evaporation step. \aj{I don't think we need to give this apology for not optimizing the evaporation ramp.}
We perform a microwave sweep to remove all Na from the magnetic trap in 1 second, which is chosen to be similar to the lifetime of NaLi with Na and longer than the thermalization time among Na ($\sim 670$ ms) and between Na and NaLi ($\sim 90$ ms) (see Methods for details). 

We compare the temperature of NaLi after the Na evaporation to that of NaLi from an identical mixture loaded to the magnetic trap but with sudden removal of Na using a resonant light pulse after the loading. 
The two temperatures are estimated by the RF-knife frequency scan described earlier (Fig.\ref{fig:atom and molecule}(b)). Evaporation of Na leads to a temperature of NaLi molecules of 0.8(1) $\mu$K, substantially lower than the temperature without the evaporation ($ 2.3(3) \;\mu$K), while the molecule number is decreased by 30\%. This corresponds to an increase in PSD of the molecular gas by a factor of $\approx 16$.

% \begin{figure}  
%     \centering  
% 	\includegraphics[width=83mm,keepaspectratio]{images/combined plot.pdf}
% 	\caption{Number decay of spin-polarized NaLi molecules in $\ket{F_{1}=5/2, F=7/2, m_{F}=7/2}$ hyperfine state are with molecules in state $\ket{F_{1}=5/2, F=5/2, m_{F}=5/2}$ is in red, with molecules in state $\ket{F_{1}=3/2, F=5/2, m_{F} =5/2}$ is in purple and with Na atoms in state $\ket{F=2,m_F=2}$ is in green.}
% 	\label{fig:mixture decay curves}
% \end{figure}

\section{Outlook}
%\wk{[I would start out by summarizing the magnetic trap can have a few advantages over optical traps:  Harmonic shape, trapping potential depends only on well-known magnetic moments, no photo-chemistry by trapping light, possible advantages in state preparation (e.g. spin flip to strong field seeking state can remove molecules in unwanted states).  The Tarbutt paper suggests that magnetic microtraps are a potential new platform for quantum simulations --- maybe a little bit far-fetched, but worth mentioning.  ]}

In summary, we have shown magnetic trapping of molecules with a factor of $10^{5}$ higher density compared to ultracold molecules previously studied in a magnetic trap. We have measured the inelastic collision rates for two selected molecular spin mixtures and a spin-stretched Na$+$NaLi mixture that serve as prototypes for future studies on state-dependent molecule-molecule and atom-molecule collisions in the magnetic trap. Quantitative analysis of molecular collisions in the magnetic trap is much simpler than in optical traps, because the magnetic trap is well-approximated by a harmonic potential whose trap frequencies are determined by molecule magnetic moments, which are typically well known. In contrast, optical traps can be highly anharmonic near the top of the trap, and ratios of trap frequencies (unless directly measured) can be difficult to determine because of the unknown molecular ac polarizability. 
%\sout{It can be challenging to experimentally measure optical trap parameters, especially for molecules in a state that is not directly detected.} 

Our collisional studies show that NaLi molecules with various collision partners have loss rates far below the universal limit. Although all collision systems are reactive, the absorption probability at close range is much smaller than one. This is well understood for the collisions in the spin-stretched Na$+$NaLi mixture \cite{son2022control, tomza2013chemical} where the quartet potential in the input channel is non-reactive, and inelastic collisions are only possible via spiflips. However, this explanation does not apply to the s-wave molecule-molecule collisions studied here. Also, a molecule-molecule Feshbach resonance has been observed for a strongly reactive input potential \cite{park2022feshbach}. These observations suggest that collisional resonances and collisional complexes should occur more generally, and  motivate more systematic studies of collisions involving NaLi and collision partners in various hyperfine states, and also for other molecules \cite{voges2022hyperfine}.  We also demonstrated sympathetic cooling of NaLi by RF evaporation of Na atoms, increasing the PSD of the molecules by a factor of $\approx 16$. This can be further optimized, but is eventually limited by the slow Na thermalization rate, which is significantly slower than the rate of elastic collisions between Na and NaLi. Using a second trap for Na that enhances the thermalization rate should allow cooling into the quantum-degenerate regime.

Our new method now allows studies of molecular collisions in a photon-free environment. Experiments probing photo-induced loss for ultracold molecular systems in optical traps have been reported using chopped optical dipole traps or repulsive box potentials made with blue-detuned light. However, even for the longest dark times and the lowest intensities, photo-induces loss could not be completely suppressed, and loss rates consistent with universal loss were observed \cite{ gregory2021molecule, ComplexesNaKNaRb, bause2021collisions}. For magnetically trapped molecules, photo-induced losses can be studied with arbitrarily small light intensities. In addition, stable microwave-induced dipoles or rotational state qubits can be achieved with technical upgrades to our apparatus. The most stable rotational qubits that have been reported are limited by the differential polarizability between two qubit states \cite{burchesky2021rotational}. In the light-free magnetic trap, we expect to produce dipoles with a longer coherence time.

% \bmhead{Data availability}
% The data that support the findings of this study are available from the corresponding author upon reasonable request.

% \bmhead{Code availability}
% The codes used to generate results are available from the corresponding author upon reasonable request.

\section{Acknowledgement}
We thank John Doyle for valuable discussions. We acknowledge support from the NSF through the Center for Ultracold Atoms and Grant No. 1506369 and from the Air Force Office of Scientific Research (MURI, Grant No. FA9550-21-1-0069). Some of the analysis was performed by W. K. at the Aspen Center for Physics, which is supported by National Science Foundation grant PHY-1607611. J. J. P. acknowledge additional support from the Samsung Scholarship. 
{\bf Author contributions:} J. J. P and Y. L carried out the experimental work. All authors contributed to the development of models, data analysis, and writing the manuscript. 

% \bmhead{Competing interest}
% The authors declare no competing interests.

%%%%%%%%%%%%%%%%%%%%%%%%%%%%%%%%%%%%%%%%%%%%%%%%%%%%%%%%%%%%%%%%%

%\newpage
\bibliography{Magtrap}

\clearpage

\newpage
\section{Methods}

	\subsection{Thermometry of molecules} 
    
    We determine the temperature of a NaLi gas in the magnetic trap from its density profile. To find this, we let NaLi molecules in the $\ket{F_{1}=5/2, F=7/2, m_{F}=7/2}$ hyperfine state undergo radio frequency (RF) induced spin-flips in a magnetic trap that has a trap bottom of 0.56 G. 
    %\aj{Why ``around''? As written it would imply you know the value with 10 mG precision. Is that right? If so, you do not need ``around.''} \jp{yes we don't need ``around''. I made some mistake in estimating the trap bottom and fixed 0.58 G to 0.56 G.}
    RF is swept from some high value ($ > 600$ kHz) to a lower value $f$ in 600 ms. This RF can induce 7 simultaneous transitions, by one quanta of $m_F$ each, from state $\ket{F_{1}=5/2, F=7/2, m_{F}=7/2}$ to $\ket{F_{1}=5/2, F=7/2, m_{F}=-7/2}$, allowing the molecules to escape from the trap. The number of molecules left in the magnetic trap is determined by the final RF value (RF-knife frequency). The maximum possible energy of the molecules is $A=\frac{7}{2} h (f-f_0)$ where $f$ is the RF-knife frequnecy and $f_0$ is the transition frequency from $\ket{F_{1}=5/2, F=7/2, m_{F}=7/2}$ to $\ket{F_{1}=5/2, F=7/2, m_{F}=5/2}$ at the bottom of the magnetic trap.
    
    Near the center, the magnetic field of the Ioffe-Pritchard magnetic trap varies quadratically with the distance from the origin. The states of the particles are enumerated by a set of quantum numbers $[n_{x} , n_{y} , n_{z} ]$ in a general three-dimensional harmonic trap potential $V(x,y,z)=\frac{1}{2}m(\omega_{x}^2x^2+\omega_{y}^2y^2+\omega_{z}^2z^2)$ and the energy of a particular state is given by $\epsilon=\hbar(n_{x}\omega_{x}+n_{y}\omega_{y}+n_{z}\omega_{z})+\epsilon_{0}$, where $\epsilon_{0}=\frac{1}{2}\hbar(\omega_{x}+\omega_{y}+\omega_{z})$ is the zero-point energy in this harmonic trap. 
    For $\epsilon \gg \epsilon_{0}$, the number of states with energy between $\epsilon$ and $\epsilon+d\epsilon$ is estimated as $g(\epsilon) d\epsilon$, with $g(\epsilon)=\frac{\epsilon^2}{2\hbar^3\bar{\omega}^3}$ where $\bar{\omega}=(\omega_x \cdot \omega_y \cdot\omega_z )^{1/3}$ is the geometric mean of the trap angular frequencies. Therefore, the particle number with energy between 0 and $A$ is
    \begin{align}	
        N(A) =& N_{\text{tot}}\frac{\int_{0}^{A}{g(\epsilon) e^{-\beta\epsilon}d\epsilon}}{\int_{0}^{\infty}{g(\epsilon) e^{-\beta\epsilon}d\epsilon}} \\
    	 =& \frac{N_{\text{tot}}}{2}\left[ e^{-\beta A}\{-\beta A \cdot (\beta A+2)-2\}  +2\right] 
    	 \label{eq:temp}
    \end{align}
    where $\beta=(k_{B}T)^{-1}$ is the Boltzmann factor. 
    Fitting the data in Fig. \ref{fig:combined}(c), which is the NaLi number as a function of the RF-knife frequency $f$, to Eq. \ref{eq:temp} provides the estimate for the molecule temperature of $1.06\pm 0.18 \; \mu$K.
    
    For the temperature estimation of NaLi in the other hyperfine state ($\ket{F_{1}=5/2, F=5/2, m_{F}=5/2}$ or $\ket{F_{1}=3/2, F=5/2, m_{F}=5/2}$) of a molecular spin mixture, we assume that the average total kinetic energy and the average total potential energy are equal in the harmonic trap by the virial theorem. After molecules are transferred from the upper-stretched state to the other hyperfine state, the average potential energy of NaLi is reduced by the magnetic moment ratio $\mu_{\text{NaLi}^{*}}/\mu_{\text{NaLi}}$ momentarily, while the kinetic energy remains the same. Here, $\text{NaLi}$ represents the upper stretched state and $\text{NaLi}^{*}$ indicates the other hyperfine state. The average potential and kinetic energy redistribute to be equal. With this simple model, the temperature is estimated as $T_{\text{NaLi}^{*}} \approx T_{\text{NaLi}}( \mu_{\text{NaLi}^{*}}/\mu_{\text{NaLi}}+1)/2$.
    
% 	\begin{figure}  
% 	    \centering  
% 		\includegraphics[width=83mm,keepaspectratio]{images/potential curves.pdf}
% 		\caption{Potential curve for rf-induced evaporation of NaLi molecules with spin $F=7/2$.}
% 		\label{fig:rf knife}
% 	\end{figure}

% 	\begin{figure}  
% 	    \centering  
% 		\includegraphics[width=83mm,keepaspectratio]{images/Mol rf.pdf}
% 		\caption{The number of molecules left in the magnetic trap is determined by the final rf value.}
% 		\label{fig:rf knife}
% 	\end{figure}

\subsection{Decay models}

The differential equations that describe the decay of molecules in the upper stretched state in the presence of another type of particle $i$ are given as:
\begin{align}
    &\dot{N}_{\text{NaLi}} = -(K_{i}/V_{\text{ov}})N_{i}N_{\text{NaLi}} - (\beta_{0}/V_{\text{eff,NaLi}}) N^2_{\text{NaLi}} \label{eq:diffeqn1} \\
    &\dot{N}_{i} = -(K_{i}/V_{\text{ov}})N_{\text{NaLi}}N_{i} - (\beta_{i}/V_{\text{eff},i}) N^2_{i}, \label{eq:diffeqn2}
\end{align}
\noindent where $N_{\rm{NaLi}}$ represents the number of NaLi in the upper stretched state and $N_{i}$ represents the number of particles of type $i$ that are magnetically trapped with the molecules. Here, $V_{\text{ov}}$ is the volume of the regime where the $i$-type particles overlap with the upper stretched molecules, $V_{\text{eff},i}$ is the mean volume filled with $i$ particles, $\beta_{0}$ ($\beta_{i}$) is the two-body molecular (particle $i$) loss rate coefficient, and $K_i$ is the loss rate coefficient for the collisions of $i$ ${+}$NaLi pairs.
% ($\ket{F_1=\frac{5}{2}, F= \frac{7}{2}, m_F=\frac{7}{2}}$) \aj{I think you've used the convention enough that it should be clear NaLi refers to the upper stretched state here. Did someone have a confusion in this part? If not, I'd remove it because the repetition actually confused me. }

We solve the given differential equations for three different conditions: $N_{i}=0$, $N_{i}\approx N_{\rm{NaLi}}$, and $N_{i}\gg N_{\rm{NaLi}}$. 
The two-body loss in the single spin component ($\ket{F_1=\frac{5}{2}, F= \frac{7}{2}, m_F=\frac{7}{2}}$) molecular gas, i.e. $N_i=0$, is described by the second term of Eq. \ref{eq:diffeqn1} only. The analytical solution is given as $N_{\text{NaLi}}(t)= N_{0}\frac{1}{1+\beta_{0}N_{0}t/V_{\text{eff}}}$,
% \begin{equation}
%     N_{\text{NaLi}}(t)= N_{0}\frac{1}{1+\frac{\beta_{0}N_{0}}{V_{\text{eff}}}t}.
% \end{equation}
where $N_0$ is the initial NaLi number.
%$\alpha$ and $\tau$ are extra variables to account for change in the trap geometry when particles are transferred from the magnetic trap to an optical trap (1550 nm 1D lattice) for detection. $\alpha$ is the rate factor (ratio of loss rate immediately after the transfer and before the transfer) and $\tau$ is the effective time spent in the lattice before detection, which is about 50 ms in this case.

For collisions in two-component mixtures that we study experimentally, the decay of the NaLi number is well described by Eq.~(\ref{eq:diffeqn1}) and (\ref{eq:diffeqn2}) with the second terms approximated to zero since $\beta_0,\beta_i \ll K_i$. With this approximation, the analytic solution becomes: 
\begin{align}
 & N_{\text{NaLi}}(t) = \frac{D}{Ce^{D\Gamma t} - 1} \label{eq:mixture 1}\\
  & N_{i}(t) = -\frac{D}{\frac{1}{C}e^{-D\Gamma t} - 1} \label{eq:mixture 2}
\end{align}
where $\Gamma=K_i N_i(0)/V_{\text{ov}}$ is the loss rate, $C = N_{i}(0)/N_{\text{NaLi}}(0)$, and $D = N_{i}(0){-}N_{\text{NaLi}}(0)$. 
% \begin{equation}
%   N_{\text{NaLi},\ket{\frac{5}{2},\frac{7}{2},\frac{7}{2}}}(t) = \frac{\text{D}}{\text{C}e^{\frac{\text{D}K_{i}}{V_{\text{ov}}}(t+\alpha \tau)} - 1}
% \end{equation}
% \\
% When $D \approx N_{i}(0)$, $N_{i}$ is approximately constant, and we find the decay curve of molecules fits well to a simple exponential function.
In a regime where $N_i(0)\gg N_{\text{NaLi}}(0)$ in a two-component mixture, $N_i(t)$ can be approximated to $N_i(0)$ throughout and Eq. \ref{eq:mixture 1} is reduced to a simple exponential decay, $ N_{\text{NaLi}}(t) = N_{0}e^{-K_{i}t/V_{\text{ov}}}$. In the experiment, the Na atom number was more than a factor of 10 larger than the NaLi molecule number, so the decay of NaLi is well described by the exponential function. 
% \begin{equation}
%     N_{\text{NaLi},\ket{\frac{5}{2},\frac{7}{2},\frac{7}{2}}}(t)= N_{0}e^{-\frac{K_{\text{Na}}}{V_{\text{ov}}}(t+\alpha \tau)}
% \end{equation}
% the two-body loss terms in Eq.~(\ref{eq:diffeqn1}) can be ignored and the Na number can be treated as constant
% The initial number of Na atoms is an order of magnitude larger than the number of molecule, and the decrease in the number of Na atoms is negligible. Molecular two-body loss caused by p-wave collisions is much slower compared to loss due to collisions with denser Na atoms.

Now we discuss the volumes $V_{\text{eff}}$ and $V_{\text{ov}}$ in Eqs.~(\ref{eq:diffeqn1}) and (\ref{eq:diffeqn2}). Assuming a harmonic trap, one obtains: 
%The total collision rate is given by $\Gamma_{\text{coll}}=\sigma_{\text{el}}v_{\rm rel}I$ where $\sigma_{\rm el}$ is the elastic scattering cross-section, $v_{\rm rel}$ is the mean relative velocity, and $I$ is the overlap density of the mixture \cite{mixtureCrosssec}:	
\begin{align*}	
%	v_{\rm rel} &= \sqrt{\frac{8k_B}{\pi} \bigg(\frac{T_{1}}{m_{1}} + \frac{T_{2}}{m_{2}}  \bigg)} \\
    V_{\text{eff},i} &= \bar{\omega}_{i}^{-3}(4\pi k_B T_{i}/m_{i})^{3/2} \\
    V_{\text{ov}} &{=} \frac{N_{i}N_{\text{NaLi}}}{\int dV n_{i}n_{\text{NaLi}}} \\
     & = \bar{\omega}_{\text{Na}}^{-3}\! \left[\!\left(\frac{2\pi k_B T_{i}}{m_{i}}\right) \left(\frac{\mu_{\text{Na}}}{\mu_{i}} + \frac{\mu_{\text{Na}}}{\mu_{2}}\frac{T_{\text{NaLi}}}{T_{i}} \right) \!\right]^{\frac{3}{2}} ,
\end{align*}
where the geometric mean of the sodium trap angular frequencies, $\bar{\omega}_{\text{Na}} {=} (\omega_x\omega_y\omega_z)^{1/3}{=} 2\pi {\times} (282 {\cdot} 282 {\cdot} 14.8)^{1/3} \; \rm{Hz} \approx 2\pi \times 106 \; \rm{Hz}$.
%\aj{Two questions: Do you really know these frequencies for $1\%$ precision or better? Why are you making me do the arithmetic here? Can we have an $\approx 106\;{\rm Hz}$? (or 110 Hz, if your uncertainties are more like 5-10\%.)}. \jp{I think we do know the frequencies for better than 1\% precision. I added $\approx 105$ Hz so you don't do the arithmetic again that you already did.} 
Here, $n_i$ is the density, $T_i$ is the temperature, $m_i$ is the mass and $\mu_i$ is the magnetic moment of an $i$ particle. The geometric mean of the trap angular frequencies for NaLi in the upper stretched state, $\bar{\omega}_{\text{NaLi}} {=} \bar{\omega}_{\text{Na}}\cdot (m_{\text{Na}}/m_{\text{NaLi}}) \cdot (\mu_{\text{NaLi}}/\mu_{\text{Na}})$, where the mass ratio, $m_{\text{NaLi}}/m_{\text{Na}} \approx 29/23$ and the magnetic moment ratio, $\mu_{\text{NaLi}}/\mu_{\text{Na}} \approx 2$. 

\subsection{Inelastic collision and thermalization rates}

To investigate the limiting factor for sympathetic cooling, we compare the three relevant time scales in a Na$+$NaLi mixture: the inelastic collision and thermalization rates between Na and NaLi, and the thermalization rate among Na atoms. The experiment was done with about $33 \times 10^{3}$ NaLi and $420 \times 10^{3} $ Na at a temperature of $\approx 2 \mu$K in a magnetic trap.
%with the geometric mean of the trap frequencies for Na, $\bar{\omega}_{\text{Na}} {=} (\omega_x\omega_y\omega_z)^{1/3}{=} 2\pi {\times} (281 {\cdot} 281 {\cdot} 14.8)^{1/3}$ Hz 
%\aj{You've already given the geometric mean once before. At most, just give a value here, not the formula.} \jp{Agreed!} 
The initial densities of Na and NaLi are $1.25 \times 10^{11} \; \rm{cm}^{-3}$ and $3.3 \times 10^{10} \; \rm{cm}^{-3}$, respectively. The initial inelastic collision rate is experimentally measured to be $\Gamma_{\rm{inel}} \approx 1.2\; \rm{s}^{-1}$ as shown in Fig. \ref{fig:atom and molecule}(a).

Next, we estimate the thermalization rate between Na and NaLi. In a mass-imbalanced system, the factor $3/\xi$ quantifies the approximate average number of collisions per particle required for thermalization, where $\xi =  4m_{\text{Na}}m_{\text{NaLi}}/(m_{\text{Na}}+m_{\text{NaLi}})^2 \approx 0.987$ \cite{LiCsmixture, NaLiSympCool}. Thus, the relation between the thermalization rate and the elastic scattering rate is given by $\Gamma_{\text{th}} \approx  \Gamma_{\text{el}}/(3/\xi)$. In our system, where the particle number is largely imbalanced, we can write the thermalization rate as

\begin{equation}
    \Gamma_{\rm th} \approx \frac{(N_{\text{Na}}/V_{\text{ov}})\sigma_{\rm el}v_{\rm rel}}{3/\xi},    
\end{equation}
where $N_{\rm{Na}}/V_{\rm{ov}}$ is the average density of Na atoms seen by NaLi molecules, $\sigma_{\rm el}$ is the elastic scattering cross-section and the relative velocity $v_{\rm{rel}} = \sqrt{\frac{8k_B}{\pi} \left(\frac{T_{\text{Na}}}{m_{\text{Na}}} + \frac{T_{\text{NaLi}}}{m_{\text{NaLi}}}  \right)}$. The $s$-wave elastic scattering cross-section between Na and NaLi is given by $\sigma_{\rm{el}} \approx 4\pi a^2$, where $a$ is the scattering length. Using an approximate value for the scattering length $a = 263(66)a_0$, where $a_0$ is the Bohr radius \cite{NaLiSympCool}, the thermalization rate is estimated to be $\approx 11 \; \rm{s}^{-1}$ .

Similarly, the thermalization rate among Na is given as $\Tilde{\Gamma}_{\rm th} = \frac{(N_{\text{Na}}/V_{\text{eff,Na}})\Tilde{\sigma}_{\rm el} \Tilde{v}_{\rm rel}}{3} \approx 1.5 \; \rm{s}^{-1}$, where the $s$-wave elastic scattering cross-section between Na atoms $\Tilde{\sigma}_{\rm{el}} \approx 8\pi \Tilde{a}^2$, where the scattering length between Na atoms $\Tilde{a}=85(3)a_{0}$ \cite{tiesinga1996spectroscopic}. The relative velocity $\Tilde{v}_{\rm{rel}} = \sqrt{\frac{16k_B}{\pi} \frac{T_{\text{Na}}}{m_{\text{Na}}} }$. We see that $\Gamma_{\rm inel} \oldsim \Tilde{\Gamma}_{\rm th} \ll\Gamma_{\rm th}$, and the sympathetic cooling of NaLi is limited by the slow Na thermalization rate.

%Given the measured thermalization rat}e, we obtain the elastic scattering cross-section between a molecule and an atom, $\sigma_{\rm el} \approx (3/\xi)(N_{\text{NaLi}}/v_{\rm rel}I)\Gamma_{\rm th}= 2.4(6) \times 10^{-11} {\rm cm}^{2}$ and the corresponding scattering length, $a = 263(66)a_0$ where $a_0$ is the Bohr radius. 

%The initial overlap density is $1.9 \times 10^{11} \; \rm{cm}^{-3}$
%Initial Na density $1.25 \times 10^{11} \; \rm{cm}^{-3}$
%Initial NaLi density $3.3 \times 10^{10} \; \rm{cm}^{-3}$

%Na-NaLi inelastic collision rate: 1.2 $s^{-1}$ 

\end{document}